%Paper: gr-qc/9405048
%From: LOUSTO@IFAE.ES
%Date: Sun, 22 May 1994 20:32:34 GMT+0100
%Date (revised): Wed, 25 May 1994 13:39:33 GMT+0100

\font\gross=cmbx10  scaled\magstep2
\font\mittel=cmbx10 scaled\magstep1
%%%------------ Mathematical symbols ------
\def\gsim{\mathrel{\raise.3ex\hbox{$>$\kern- .75em
                      \lower1ex\hbox{$\sim$}}}}
\def\lsim{\mathrel{\raise.3ex\hbox{$<$\kern-.75em
                      \lower1ex\hbox{$\sim$}}}}
\def\square{\kern 1pt
\vbox{\hrule height 0.6pt\hbox{\vrule width 0.6pt \hskip 3pt
\vbox{\vskip 6pt}\hskip 3pt\vrule width 0.6pt} \hrule height 0.6pt}\kern 1pt }

\def\sla{\raise.15ex\hbox{$/$}\kern-.72em}

%%%---------------------------------------------------

\parskip=\medskipamount
\overfullrule=0pt
\raggedbottom
\def\normalparindent{24pt}
\newif\ifdraft \draftfalse

\nopagenumbers
\footline={\ifnum\pageno=1 {\ifdraft
{\hfil\rm Draft \number\day -\number\month -\number\year}
\else{\hfil}\fi}
\else{\hfil\rm\folio\hfil}\fi}
\def\endpage{\vfill\eject}
\def\beginlinemode{\endmode\begingroup\parskip=0pt
\obeylines\def\\{\par}\def\endmode{\par\endgroup}}
\def\beginparmode{\endmode\begingroup \def\endmode{\par\endgroup}}
\let\endmode=\par
\def\raggedcenter{%\leftskip=4em plus 12em \rightskip=\leftskip
                  \leftskip=2em plus 6em \rightskip=\leftskip
                  \parindent=0pt \parfillskip=0pt \spaceskip=.3333em
                  \xspaceskip=.5em\pretolerance=9999 \tolerance=9999
                  \hyphenpenalty=9999 \exhyphenpenalty=9999 }
\def\\{\cr}
\let\rawfootnote=\footnote\def\footnote#1#2{{\parindent=0pt\parskip=0pt
        \rawfootnote{#1}{#2\hfill\vrule height 0pt depth 6pt width 0pt}}}
\def\title{\null\vskip 3pt plus 0.2fill\beginlinemode\raggedcenter\gross}
\def\author{\vskip 3pt plus 0.2fill \beginlinemode\raggedcenter}
\def\affil{\vskip 3pt plus 0.1fill\beginlinemode\raggedcenter\it}
\def\abstract{\vskip 3pt plus 0.3fill \beginparmode{\noindent
{\mittel Abstract}:~}  }
\def\endtitlepage{\endpage\body}
\def\body{\beginparmode\parindent=\normalparindent}
\def\head#1{\par\goodbreak{\immediate\write16{#1}
      \vskip 0.4cm{\noindent\gross #1}\par}\nobreak\nobreak\nobreak\nobreak}

%--------  Definition of Journal Styles --------
\def\finalcite{\citeall\ref\citeall\Ref}
\newif\ifannpstyle
\newif\ifprdstyle
\newif\ifplbstyle
\newif\ifwsstyle

\gdef\refto#1{\ifprdstyle  $^{\[#1] }$ \else
              \ifwsstyle$^{\[#1]}$  \else
              \ifannpstyle $~[\[#1] ]$ \else
              \ifplbstyle  $~[\[#1] ]$ \else
                                         $^{[\[#1] ]}$\fi\fi\fi\fi}
\gdef\refis#1{\ifprdstyle \item{~$^{#1}$}\else
              \ifwsstyle \item{#1.} \else
              \ifplbstyle\item{~[#1]} \else
              \ifannpstyle \item{#1.} \else
                              \item{#1.\ }\fi\fi\fi\fi }
\gdef\journal#1,#2,#3,#4.{
           \ifprdstyle {#1~}{\bf #2}, #3 (#4).\else
           \ifwsstyle {\it #1~}{\bf #2~} (#4) #3.\else
           \ifplbstyle {#1~}{#2~} (#4) #3.\else
           \ifannpstyle {\sl #1~}{\bf #2~} (#4), #3.\else
                       {\sl #1~}{\bf #2}, #3 (#4)\fi\fi\fi\fi}

%----------------------------------------------
\def\ref#1{Ref.~#1}
\def\Ref#1{Ref.~#1}
\def\cite#1{{#1}}\def\[#1]{\cite{#1}}

\def\eq#1{Eq.~\(#1)}\def\eqs#1{Eqs.~\(#1)}

\def\(#1){(\call{#1})}
\def\call#1{{#1}}\def\taghead#1{{#1}}

\def\references{\head{References}\beginparmode\frenchspacing\parskip=0pt}
\def\endreferences{\body}
\def\endit{\endmode\vfill\supereject}\let\endpaper=\endit
%%%%%%--------Journal  Abbreviations  -------------

%%-------------- Abbreviated Adresses  --------------------

\def\kndir{Fakult\"at f\"ur Physik der Universit\"at Konstanz,\\
Postfach 5560 M 674, D-78434 Konstanz, Germany.}%\\E-mail: phlousto@dknkurz1}

\def\barce{IFAE - Grupo de F\'\i sica Te\'orica,\\
Universidad Aut\'onoma de Barcelona,\\
E-08193 Bellaterra (Barcelona), Spain.\\ E-mail: lousto@ifae.es}

\def\iafedir{Permanent address: Instituto de Astronom\'\i a y F\'\i sica del
Espacio, Casilla de Correo 67 -\\ Sucursal 28, 1428 Buenos Aires, Argentina.
E-mail: lousto@iafe.edu.ar}

\def\mangos{This work was partially supported by the Directorate General for
Science, Research and Development of the Commission of the European
Communities.}

\def\dos{ C.O.L was also supported by the Alexander von Humboldt Foundation
and by the Direcci\'on General de
Investigaci\'on Cient\'\i fica y T\'ecnica of the Ministerio de Educaci\'on
y Ciencia de Espa\~na and CICYT AEN93-0474.}

%%----------------------------------------------------------
%%%%%%%%%% REFORDER + EQNORDER
\catcode`@=11
\newcount\r@fcount \r@fcount=0\newcount\r@fcurr
\immediate\newwrite\reffile\newif\ifr@ffile\r@ffilefalse
\def\w@rnwrite#1{\ifr@ffile\immediate\write\reffile{#1}\fi\message{#1}}
\def\writer@f#1>>{}
\def\referencefile{\r@ffiletrue\immediate\openout\reffile=\jobname.ref%
  \def\writer@f##1>>{\ifr@ffile\immediate\write\reffile%
    {\noexpand\refis{##1} = \csname r@fnum##1\endcsname = %
     \expandafter\expandafter\expandafter\strip@t\expandafter%
     \meaning\csname r@ftext\csname r@fnum##1\endcsname\endcsname}\fi}%
  \def\strip@t##1>>{}}

\def\citeall#1{\xdef#1##1{#1{\noexpand\cite{##1}}}}
\def\cite#1{\each@rg\citer@nge{#1}}
\def\each@rg#1#2{{\let\thecsname=#1\expandafter\first@rg#2,\end,}}
\def\first@rg#1,{\thecsname{#1}\apply@rg}
\def\apply@rg#1,{\ifx\end#1\let\next=\relax%
\else,\thecsname{#1}\let\next=\apply@rg\fi\next}%
\def\citer@nge#1{\citedor@nge#1-\end-}
\def\citer@ngeat#1\end-{#1}
\def\citedor@nge#1-#2-{\ifx\end#2\r@featspace#1
  \else\citel@@p{#1}{#2}\citer@ngeat\fi}
\def\citel@@p#1#2{\ifnum#1>#2{\errmessage{Reference range #1-#2\space is bad.}
    \errhelp{If you cite a series of references by the notation M-N, then M and
    N must be integers, and N must be greater than or equal to M.}}\else%
{\count0=#1\count1=#2\advance\count1 by1\relax\expandafter\r@fcite\the\count0,%
  \loop\advance\count0 by1\relax%         Loop from M to N
    \ifnum\count0<\count1,\expandafter\r@fcite\the\count0,%
  \repeat}\fi}
\def\r@featspace#1#2 {\r@fcite#1#2,}    \def\r@fcite#1,{\ifuncit@d{#1}
    \expandafter\gdef\csname r@ftext\number\r@fcount\endcsname%
    {\message{Reference #1 to be supplied.}\writer@f#1>>#1 to be supplied.\par
     }\fi\csname r@fnum#1\endcsname}
\def\ifuncit@d#1{\expandafter\ifx\csname r@fnum#1\endcsname\relax%
\global\advance\r@fcount by1%
\expandafter\xdef\csname r@fnum#1\endcsname{\number\r@fcount}}
\let\r@fis=\refis   \def\refis#1#2#3\par{\ifuncit@d{#1}%
    \w@rnwrite{Reference #1=\number\r@fcount\space is not cited up to now.}\fi%
  \expandafter\gdef\csname r@ftext\csname r@fnum#1\endcsname\endcsname%
  {\writer@f#1>>#2#3\par}}
\def\r@ferr{\endreferences\errmessage{I was expecting to see
\noexpand\endreferences before now;  I have inserted it here.}}
\let\r@ferences=\references
\def\references{\r@ferences\def\endmode{\r@ferr\par\endgroup}}
\let\endr@ferences=\endreferences
\def\endreferences{\r@fcurr=0{\loop\ifnum\r@fcurr<\r@fcount
    \advance\r@fcurr by 1\relax\expandafter\r@fis\expandafter{\number\r@fcurr}%
    \csname r@ftext\number\r@fcurr\endcsname%
  \repeat}\gdef\r@ferr{}\endr@ferences}
\let\r@fend=\endpaper\gdef\endpaper{\ifr@ffile
\immediate\write16{Cross References written on []\jobname.REF.}\fi\r@fend}
\catcode`@=12
\finalcite
%\citeall\refto\citeall\ref\citeall\Ref
%-----------------
\catcode`@=11
\newcount\tagnumber\tagnumber=0
\immediate\newwrite\eqnfile\newif\if@qnfile\@qnfilefalse
\def\write@qn#1{}\def\writenew@qn#1{}
\def\w@rnwrite#1{\write@qn{#1}\message{#1}}
\def\@rrwrite#1{\write@qn{#1}\errmessage{#1}}
\def\taghead#1{\gdef\t@ghead{#1}\global\tagnumber=0}
\def\t@ghead{}\expandafter\def\csname @qnnum-3\endcsname
  {{\t@ghead\advance\tagnumber by -3\relax\number\tagnumber}}
\expandafter\def\csname @qnnum-2\endcsname
  {{\t@ghead\advance\tagnumber by -2\relax\number\tagnumber}}
\expandafter\def\csname @qnnum-1\endcsname
  {{\t@ghead\advance\tagnumber by -1\relax\number\tagnumber}}
\expandafter\def\csname @qnnum0\endcsname
  {\t@ghead\number\tagnumber}
\expandafter\def\csname @qnnum+1\endcsname
  {{\t@ghead\advance\tagnumber by 1\relax\number\tagnumber}}
\expandafter\def\csname @qnnum+2\endcsname
  {{\t@ghead\advance\tagnumber by 2\relax\number\tagnumber}}
\expandafter\def\csname @qnnum+3\endcsname
  {{\t@ghead\advance\tagnumber by 3\relax\number\tagnumber}}
\def\equationfile{\@qnfiletrue\immediate\openout\eqnfile=\jobname.eqn%
  \def\write@qn##1{\if@qnfile\immediate\write\eqnfile{##1}\fi}
  \def\writenew@qn##1{\if@qnfile\immediate\write\eqnfile
    {\noexpand\tag{##1} = (\t@ghead\number\tagnumber)}\fi}}
\def\callall#1{\xdef#1##1{#1{\noexpand\call{##1}}}}
\def\call#1{\each@rg\callr@nge{#1}}
\def\each@rg#1#2{{\let\thecsname=#1\expandafter\first@rg#2,\end,}}
\def\first@rg#1,{\thecsname{#1}\apply@rg}
\def\apply@rg#1,{\ifx\end#1\let\next=\relax%
\else,\thecsname{#1}\let\next=\apply@rg\fi\next}
\def\callr@nge#1{\calldor@nge#1-\end-}\def\callr@ngeat#1\end-{#1}
\def\calldor@nge#1-#2-{\ifx\end#2\@qneatspace#1 %
  \else\calll@@p{#1}{#2}\callr@ngeat\fi}
\def\calll@@p#1#2{\ifnum#1>#2{\@rrwrite{Equation range #1-#2\space is bad.}
\errhelp{If you call a series of equations by the notation M-N, then M and
N must be integers, and N must be greater than or equal to M.}}\else%
{\count0=#1\count1=#2\advance\count1 by1\relax\expandafter\@qncall\the\count0,%
  \loop\advance\count0 by1\relax%
    \ifnum\count0<\count1,\expandafter\@qncall\the\count0,  \repeat}\fi}
\def\@qneatspace#1#2 {\@qncall#1#2,}
\def\@qncall#1,{\ifunc@lled{#1}{\def\next{#1}\ifx\next\empty\else
  \w@rnwrite{Equation number \noexpand\(>>#1<<) has not been defined yet.}
  >>#1<<\fi}\else\csname @qnnum#1\endcsname\fi}
\let\eqnono=\eqno\def\eqno(#1){\tag#1}\def\tag#1$${\eqnono(\displayt@g#1 )$$}
\def\aligntag#1\endaligntag  $${\gdef\tag##1\\{&(##1 )\cr}\eqalignno{#1\\}$$
  \gdef\tag##1$${\eqnono(\displayt@g##1 )$$}}
\def\eqalignno#1{\displ@y \tabskip\centering
  \halign to\displaywidth{\hfil$\displaystyle{##}$\tabskip\z@skip
    &$\displaystyle{{}##}$\hfil\tabskip\centering
    &\llap{$\displayt@gpar##$}\tabskip\z@skip\crcr
    #1\crcr}}
\def\displayt@gpar(#1){(\displayt@g#1 )}
\def\displayt@g#1 {\rm\ifunc@lled{#1}\global\advance\tagnumber by1
        {\def\next{#1}\ifx\next\empty\else\expandafter
        \xdef\csname @qnnum#1\endcsname{\t@ghead\number\tagnumber}\fi}%
  \writenew@qn{#1}\t@ghead\number\tagnumber\else
        {\edef\next{\t@ghead\number\tagnumber}%
        \expandafter\ifx\csname @qnnum#1\endcsname\next\else
        \w@rnwrite{Equation \noexpand\tag{#1} is a duplicate number.}\fi}%
  \csname @qnnum#1\endcsname\fi}
%-------------
\def\eqnoa(#1){\global\advance\tagnumber by1\multitag{#1}{a}}
\def\eqnob(#1){\multitag{#1}{b}}
\def\eqnoc(#1){\multitag{#1}{c}}
\def\eqnod(#1){\multitag{#1}{d}}
\def\multitag#1#2$${\eqnono(\multidisplayt@g{#1}{#2} )$$}
\def\multidisplayt@g#1#2 {\rm\ifunc@lled{#1}
        {\def\next{#1}\ifx\next\empty\else\expandafter
        \xdef\csname @qnnum#1\endcsname{\t@ghead\number\tagnumber b}\fi}%
  \writenew@qn{#1}\t@ghead\number\tagnumber #2\else
        {\edef\next{\t@ghead\number\tagnumber #2}%
        \expandafter\ifx\csname @qnnum#1\endcsname\next\else
    \w@rnwrite{Equation \noexpand\multitag{#1}{#2} is a duplicate number.}\fi}%
  \csname @qnnum#1\endcsname\fi}
%-----------------
\def\ifunc@lled#1{\expandafter\ifx\csname @qnnum#1\endcsname\relax}
\let\@qnend=\end\gdef\end{\if@qnfile
\immediate\write16{Equation numbers written on []\jobname.EQN.}\fi\@qnend}
%\catcode`@=12

%%%%%%%%%%%%%%%%%%%%%%%%%%%%%%%%%%%%%%%%%%%%%%%%%%%%%%%%%%%%%%%%%%%%%%%%%%%%

\baselineskip=14pt
\magnification=1200
\parindent=0pt
\title
The Emergence of an Effective two -

dimensional Quantum Description from the

study of Critical Phenomena in Black Holes

\author
C. O. Lousto\footnote{$^*$}{\iafedir}

\affil\barce
and
\kndir

\abstract

We study the occurrence of critical phenomena in four - dimensional,
rotating and charged black holes, derive the critical exponents and
show that they fulfill the scaling laws. Correlation functions critical
exponents and Renormalization Group considerations assign an effective
(spatial) dimension, $d=2$, to the system.
The two - dimensional Gaussian approximation to critical systems is shown
to reproduce all the black hole's critical exponents. Higher order corrections
(which are always relevant) are discussed. Identifying the two - dimensional
surface with the event horizon and noting that generalization of scaling
leads to conformal invariance and then to string theory, we arrive to
't Hooft's string interpretation of black holes. From this, a model for
dealing with a coarse grained black hole quantization is proposed.
We also give simple arguments that lead to a rough
quantization of the black hole
mass in units of the Planck mass, i. e. $M\simeq{1\over\sqrt{2}}M_{pl}
\sqrt{l}$ with a $l$ positive integer and then, from this result, to
the proportionality between quantum entropy and area.

\endtitlepage
\head{1. Introduction}
\parindent=15pt

The scaling of critical phenomena\refto{19,18,A84,B92}
applies to a great variety of thermodynamical systems. Those ranging from
the internal structure of elementary particles to ferroelectricity and
turbulent fluid flow, passing through superconductivity and superfluidity.
The scaling is found to hold (within experimental error) in almost every case.
The renormalization group approach \refto{18, WK74} use the scaling hypothesis
and provides a sound mathematical foundation to the concept of universality.
On the other hand black hole dynamics is governed by analogues of
the ordinary four laws of thermodynamics\refto{1,9,W92}. This
two facts lead us to conjecture that black holes also obey the scaling
laws or fourth law of thermodynamics\refto{L93,L94}:

Let us suppose that a rotating charged black hole is held in equilibrium at
some temperature $T$, with a surrounding heat bath. If we consider a small,
reversible transfer of energy between the hole and its environment; this
absorption will be isotropic, and will occur in such a way that the angular
momentum $J$ and charge $Q$ remain unchanged, on the average. The full
thermal capacity (not per unit mass) corresponding to this energy transfer can
be computed by eliminating $M$ (the total mass of the black hole)
between the equations for the temperature
and the area of the black hole, and differentiating the entropy, $S$,
keeping $J$ and $Q$ constant\refto{L93},
$$
C_{J,Q}=T{\partial S\over\partial T}\biggr\vert_{J,Q}={MTS^3\over\pi J^2+
{\pi\over4}Q^4-T^2S^3}~.\eqno(25)
$$

This heat capacity goes from negative values for a Schwarzschild black hole,
$C_{Sch}=-M/T$, to positive values for a nearly extreme Kerr - Newman black
hole, $C_{EKN}\sim \sqrt{M^4-J^2-M^2Q^2} \to 0^+$. Thus, $C_{J,Q}$ has changed
sign at some value of
$J$ and $Q$ in between. In fact, the heat capacity passes from negative to
positive values through an infinite discontinuity. This feature has lead
Davies\refto{4} to classify the phenomenon at the critical values of
$J$ and $Q$ as a second order phase transition. The critical
values $J_c$ and $Q_c$
at which the transition occurs are obtained by making to vanish the denominator
on the right hand side of eq \(25). We can then define the following
parametrization,
$$
J_c^2={j\over8\pi}M^4~~~{\rm and}~~~Q_c^2={q\over8\pi}M^2~.
$$
Eliminating $S$ and $T$ in eq \(25) by use of the expressions for the
temperature and entropy of a black hole\refto{L93},
the infinite discontinuity in $C_{J,Q}$ takes place at\refto{4}
$$
j_{JQ}^2+6j_{JQ}+4q_{JQ}=3~.\eqno(27)
$$
For an uncharged, i.e., Kerr, hole, $q_{JQ}=0$. Thus, $j_{JQ}=2\sqrt{3}-3$.
Then we have
$$
\Omega_c={\sqrt{2\sqrt{3}-3}\over4\sqrt{3}-3}T_c\cong0.233T_c~.\eqno(28)
$$
where $\Omega$ is the angular velocity of the event horizon.\par\noindent
While for a non rotating, i.e., Reissner - Nordstr\"om, hole, $j_{JQ}=0$. Thus,
$q_{JQ}=3/4$. And the critical value of the electric potential is given by
$$
\Phi_c={1\over\sqrt{3}}~,\eqno(29)
$$
independent of the other parameters of the black hole such as its mass or
charge.

It can also be shown\refto{26} that
the four isothermal compressibilities, $K^{-1}$,
are divergent as their corresponding heat capacities.
For example,
$$
K_{T,Q}^{-1}=J{\partial\Omega\over\partial J}\biggr\vert_{T,Q}\sim
{\pi(2\Phi Q-M)(1-4\pi TM)\over S^2[1-12\pi TM+4\pi^2T^2(6M^2+Q^2)]}~,
\eqno(30)
$$
diverges as $C_{J,Q}$.
Also
$K_{T,J}^{-1}=C_{J,Q}(\partial\Phi/\partial Q\big\vert_{S,J})/C_{J,\Phi}$
diverges as $C_{J,Q}$ on the singular
segment given by eq \(27).

By use of the expressions for the temperature and entropy of
black holes\refto{L93},  the heat
capacity $C_{JQ}$ can be expressed as \refto{26}
$$
C_{JQ}={4\pi TSM\over 1-8\pi TM-4\pi ST^2}\sim {1\over T-T_c} ~,\eqno(31)
$$
where the critical temperature is given by
$T_c^{JQ}= \left\{2\pi M[3+\sqrt{3-q_{JQ}}]\right\}^{-1} ~,$
and $q_{JQ}$ is given by the critical curve \eq{27}.

We can obtain the  first two critical exponents (that characterizes the
approach to the divergence at $J,Q$ and $T$ fixed, respectively\refto{L93})
directly by inspection of eq \(31):
$$
\alpha=1~~~,~~~\varphi=1~~.\eqno(35)
$$
Analogously, from the expression for the compressibility, eq \(30)
(that diverges as $C_{J,Q}$), we obtain the critical exponents corresponding
to the
approach to the divergence at $J,Q$ and $T$ fixed, respectively)
$$
\gamma=1~~~,~~~1-\delta^{-1}=1~\Rightarrow~\delta^{-1}\to0~~.\eqno(36)
$$

To obtain the critical exponents corresponding to the equation of state and
entropy, we choose a path either along a critical isotherm or at constant
angular momentum $J=J_c$ or constant charge $Q=Q_c$.
However, in this case the black hole equations of state just
reproduce the critical curves (such as eqs \(28)-\(29), and others
deduced from them).
In this case, we can formally assign a zero power
corresponding to the critical exponents associated to $\Omega$ and $S$
respectively\refto{L93}:
$$
\beta\to0~~~,~~~\delta^{-1}\to0~~,
$$
$$
1-\alpha=0~~~,~~~\psi\to0~~.\eqno(37)
$$
One can easily check that the set of critical values given by eqs \(35)-\(37)
satisfy the scaling laws\refto{19} (with $\beta\delta=1$):
$$
\alpha+2\beta+\gamma=2~~, ~~~\alpha+\beta(\delta +1)=2 ~~,
$$
$$
\gamma(\delta +1)=(2-\alpha)(\delta-1)~~,
{}~~\gamma=\beta(\delta -1)~~, \eqno(24)
$$
$$
(2-\alpha)(\delta\psi-1)+1=(1-\alpha)\delta~~,
{}~~\varphi+2\psi-\delta^{-1}=1~~.
$$

Other five heat capacities can be computed, of which $C_{\Omega,Q}$ and
$C_{J, \Phi}$ exhibit also a singular behavior. The remaining $C_{\Phi,Q}=
C_{J,\Omega}$ and $C_{\Omega,\Phi}$ being regular functions in the allowed
set of values of the parameters\refto{26}. Heat capacities and isothermal
compressibilities at fixed $(\Omega,Q)$ and $(J,\Phi)$ give the same
critical exponents as in the previous case where we held $(J,Q)$ constant.
This result can in fact be understood as a realization of the
{\it Universality hypothesis}: For a continuous
phase transition the static critical exponents depend only on the following
three properties:

\noindent a) the dimensionality of the system, $d$.

\noindent b) the internal symmetry dimensionality of the order parameters, $D$.

\noindent c) whether the forces are of short or long range.

The critical curves for the three
cases studied\refto{L93} are all of them different, but
the critical exponents, according to the above mentioned hypothesis, are
the same within each class as specified after eq \(24).
We also observe that the equality between the primed ($T\to T_c^-$)
and unprimed ($T\to T_c^+$) critical exponents
is trivially verified in each one of the three
transitions studied.

The lack of qualitative change in the properties of the black hole can
be understood as in analogy to what happens in the case of a liquid - vapor
system; where near criticality no qualitative distinction can be made between
phases. Note that in this case there is not such thing as a latent heat
\refto{32}(since $M$ remains continuous through the transition), as it happens
in magnetic critical transitions.
Besides, it can be seen that the critical transitions occur when we cold down
the black hole with respect to the corresponding Schwarzschild temperature,
$T_S=1/(8\pi M)$, by increasing its charge or angular momentum at fixed
total mass.
Further, we have seen how black holes fulfil the scaling laws and universality
hypothesis, both characteristics of critical phase transitions.

It worth noting\refto{32} that although this phase transition
does not affect the internal state of the system it is physically important as
it indicates  the transition from a region ($C_{JQ}<0$) where only a
microcanonical ensemble is appropriate (stable equilibrium if the system is
isolated from the outside world) to a region ($C_{JQ}>0$) where a canonical
ensemble can be also used (stable equilibrium with an infinite heat bath).

The paper is organized as follows.
In section 2 we analyse two further critical exponents defined for the static
correlation functions. We find that all critical exponents correspond to
those of a gaussian model in two-dimensions. Renormalization Group arguments
are given to stablish as $d=2$ the effective dimension of the system. In
section 3 we develop the idea of a two-dimensional effective representation
for the black hole horizon as the fundamental object to quantize and make
connection with string theory in a description similar to that of 't Hooft.
We end the paper with a discussion and simple derivation, using the
two-dimensional representation of black holes, of a mass quantization and
a quantum originated entropy-area relation.

\head{2. Correlation functions, the Gaussian model and the Renormalization
Group}

Not only relations among critical exponents corresponding to thermodynamic
functions can be obtained, but also relations concerning correlation functions
exponents.

The static two - point (at distance $|\vec r|$)
connected correlations can be defined as
$$
G_c^{(2)}(|\vec r|)=<\phi(0)\cdot\phi(\vec r)>-|<\phi>|^2~~,  \eqno(a)
$$
where $\phi$ is the order parameter of the system in question and may
have, in principle, $D$ different internal components (for example, in the
Ising model, the order parameter has only one component; in a Heisenberg
system, three, and in the ${\rm He}^3$
superfluid transition as many as eighteen\refto{A84}).

Away but not far from the critical region, one can write
$$
G_c^{(2)}(r)\sim {\exp{\{-r/\xi\}}\over r^{d-2+\eta}}~~,
{}~~r~~{\rm large}~~. \eqno(b)
$$
Here $d$ is the (spatial)
dimensionality of the system, $\eta$ is a further critical
exponent and $\xi$ is the correlation length.
As one approaches the critical curve $\xi$ diverges as
$$
\xi\sim |T-T_c|^{-\nu}~~. \eqno(c)
$$
Here $\nu$ is another critical exponent.

Kadanoff\refto{K66} studied the scaling properties of the correlation
functions and found a new scaling law relating the critical exponents
$$
(2-\eta)\nu=\gamma \eqno(d)
$$
With an additional assumption about the scaling behavior of the correlation
function \refto{HS72} one obtains the hyperscaling law
$$
\nu d=2-\alpha~~. \eqno(e)
$$
Note that only here the dependence with the dimensionality of the system
appears.
By use of the Renormalization Group equations, one can show\refto{F82}
that hyperscaling holds for $d\leq 4$ but break down for $d>4$.

Now, let us consider the black hole in equilibrium with a radiation bath.
By use of the  Quantum Field Theory technics in the curved spacetime of the
black hole one can obtain an approximate expression for
the correlation function of the fluctuations of fields
in this curved background. In equilibrium, the field will be in the
Hartle - Hawking vacuum state.
The correlation function of a scalar field in the Schwarzschild background,
for large distances $r$ is given
by\refto{SCD81} ($\omega$ being the frequency of the mode in the Hartle -
Hawking state considered)
$$
G_{\omega}(r)\sim
{\omega\over 2\pi\left[\exp\left({2\pi\omega\over k_H}\right)-1\right]}
{}~~. \eqno(f)
$$
And thus, independent of the distance $r$. Here, we shall make the hypothesis
that in equilibrium gravitational correlations from black hole fluctuations
behave in a similar qualitative way to the scalar field fluctuations,
even considering charged and rotating black holes. We will propose below that
the black hole itself can be represented effectively, near criticality, by
an order parameter, $\phi$, having the same critical exponents.

{}From \eq{b} we thus conclude that
$$
d-2+\eta=0~~. \eqno(g)
$$
The correlation length can be formally defined from \refto{A84}
$$
\xi^2=-{1\over 2K_T}\left({\partial^2G(\omega)\over\partial\omega^2}\right)
_{\omega=0}~~.\eqno(h)
$$
By use of \eq{f} and since $K_T^{-1}\sim |T-T_c|^{-\gamma}$ we find that
$$
\xi^2\sim |T-T_c|^{-1} ~~~\Rightarrow~~~\nu={1\over 2}~~, \eqno(i)
$$
where we have used the definition of $\nu$, \eq{c}, and that in our case
$\gamma=1$.

We see that our two new critical exponents  (Eqs \(g) and \(i))
take values that fulfil the scaling relation \(d) only if the dimension of the
system is $d=2$. In this case, the hyperscaling relation \(e) is also
 satisfied, as expected for $d<4$, and we have
$$
\eta=0~~. \eqno(j)
$$
This is, in fact, the first hint that our system behaves as an effective
two - dimensional one.

Additional insight can be gained by comparison to the Gaussian Model.
This model can be described in its continuous version by the partition
function\refto{B92}
$$
{\cal Z}_G(J)={\cal N}\int{{\cal D}\phi\exp\left\{-\int{d^dx
\left[{1\over 2}c^2|\vec\nabla\phi|^2+{1\over 2}\mu\phi^2-J
\phi\right]}\right\}}~~. \eqno(k)
$$
The hamiltonian appearing in the exponential can be seen as a truncation
of orders $\phi^4$ or higher in a Ginzburg - Landau model. The Gaussian model
was originally studied\refto{BK52} for a discrete spin variable. It has the
advantage of being exactly soluble and it presents a critical point with
critical exponents (for a one - component field $\phi$) given by \refto{F82}
$$
\alpha=2-d/2~~~, ~~~\beta=(d-2)/4~~~,~~~ \gamma=1~~~,
$$
$$
\delta={d+2\over d-2}~~~,~~~\eta=0~~~,~~~\nu={1\over 2}~~.\eqno(l)
$$
It is worth to remark here that all this critical exponents can be made to take
exactly the same values as for the black hole case, i.e. \eqs{35}-\(37), for
$d=2$. Thus, $d=2$, appears here as the effective spatial dimensionality
of black holes near critical conditions.

The Gaussian model is not fully satisfactory because it has no ``ordered"
phase. The integral in \eq{k} diverges for $T<T_c$ and thus one must
include higher order terms (e.g. $\phi^4$) to stabilize this integral.
It is interesting to note here that black holes themselves pass
through the critical curve (at $T=T_c$) from a region of canonical instability
to a region of canonical stability as one lowers their temperature (see fig. 1
of ref.[\cite{L93}]).

One might think that the resulting effective dimension of the system, $d=2$,
relays only on comparison to the Gaussian model and that other possibilities
are still open. To explore this possibility we can recall some results
from the Renormalization Group Theory.
Let us suppose that our effective Hamiltonian contains terms of order higher
than in the Gaussian model.
We then can write
$$
H_{eff}(\phi)={1\over 2} c^2|\nabla\phi|^2+{1\over 2}\mu\phi^2+{\lambda\over4!}
\phi^4+b\phi^2|\nabla\phi|^2+.... \eqno(m)
$$
The scaling properties of the the additional operators, with $n$ powers of
$\phi$ and $p$ derivatives, can be studied in terms of the sign of
$$\Delta=n-p-{1\over2}d(n-2)~.\eqno(d')$$
If $\Delta$ is positive / negative the operator is relevant / irrelevant
\refto{B92}.

Thus, if the dimensionality of the system where larger than or equal to four,
the Renormalization Group analysis tells us \refto{WK74, B92} that the
operators we have added
to the gaussian hamiltonian are ``irrelevant" in the sense that they do not
contribute to modify the critical exponents, which will be those of the
Gaussian
model or the mean field (Landau) theory.
Thus, no matching with the black hole results can be made for $d\geq 4$ models.
There is still the possibility of having $d=3$, as is the case of most
realistic system, e.g. those studied in the Laboratory. In $d=3$, $\phi^4$
becomes a relevant operator. One can make a perturbation theory based on the
gaussian part of the hamiltonian and obtain a set of critical exponents
\refto{B92} that fit very well with Lab
experiments but are not those of black holes.
Thus, we are left with $d=2$ (since for $d<2$ no critical phenomena takes
place). The problem here is that {\it all} operators of the form $\phi^{2n}$
and $|\nabla\phi|^2\phi^{2n}$ are relevant and thus will modify the critical
exponents. For instance, if we take the $\phi^4$ term in \eq{m},
we will have \refto{A84} to a good approximation (practically the same values
as in the Ising model) the following critical exponents in $d=2$,
$$
\alpha=0~~~,~~~\beta=1/8~~~,~~~\gamma=7/4~~~,$$
$$\delta=15~~~,~~~\eta=1/4~~~,~~~\nu=1~~.\eqno(n)$$

Since all operators are relevant, we expect this theory to be renormalizable.
In fact, we know that field theory (as well as gravity) in two dimensions
is asymptotically free in the ultraviolet allowing us to built up
a finite quantum field theory.

We can now conclude that the first order approximation to Quantum effects
in black holes correspond to the Gaussian approximation. Let us recall that
the path integral formulation of the Hawking \refto{HH76} radiation and black
hole gravitational thermodynamics relies on the stationary phase approximation
to obtain a convergent Gaussian integral. The next order approximation should
include  back reaction and self - interaction effects as well as higher order
quantum corrections.
In fact, whatever would be the final form of the quantum theory of gravity,
we can assume that the Kerr metric should be a classical solution to the
vacuum field equations. The critical exponents of this black hole solution
will then be those given in \eqs{35}-\(37). By applying the universality
hypothesis this exponents will be the same for the full family of black
hole solutions to the theory.
We can thus conjecture that the critical phenomena in black hole will survive
to higher order corrections; that the scaling laws will continue to hold, but
the critical exponents that will fulfil this laws, when quantum higher order
corrections are taken into account, will be different from those given by
\eqs{35}-\(37), and in particular, to the next quantum order to Hawking
radiation approximation they will take \eq{n} values.

\head{3. The black hole horizon as a quantum critical system}

Now that we have established that the dimensionality of the system is $d=2$,
it remains to identify this two - dimensional surface. A natural choice is the
horizon of the black hole\footnote{$^\dagger$}{Finite size effects are
not expected to affect
the scaling properties derived in the thermodynamic limit \refto{A84}.}
(or better, a slightly shifted outwards 2+1 hypersurface\refto{STU93}).
One observer far away from the black hole sees all the matter of the
collapsing body that will form it to accumulate on the two - dimensional
horizon forming some kind of ``membrane" \refto{TPM86} or subtle ``skin"
\refto{W92}.

By analogy to the models for spin systems able to suffer critical transitions
we can think of the event horizon as having only a finite (and eventually
discrete) number of degrees of freedom at every (lattice) cell of
Planckian dimensions.
We know that if there is a continuous internal symmetry  in the order
parameter, no long range order, or broken symmetry, will occur in two space
dimensions. If the symmetry is discrete it is possible (e.g. Ising model).

It is interesting to compare our approach to black hole quantization with
that of 't Hooft \refto{T90,TH92} since several points in common can be drawn.
In this approach to the problem of black hole quantization it is postulated
the existence of an S - matrix to describe the evaporation process.
This hypothesis seems to be supported by new evidence revealing that
the stimulated emission in the  Hawking radiation might play an important
role in solving the loss of quantum information/coherence paradox\refto{LM93}.
The horizon shift produced by light particles going out or coming into the
horizon is an essential ingredient in the construction of the S - matrix.
Its elements are given by
$$
<p^{out}(\Omega)| p^{in}(\Omega)>={\cal N}\exp{\left\{i\int{\int{p^{out}
(\Omega)f(\Omega-\Omega')p^{in}(\Omega')d^2\Omega d^2\Omega'}}\right\}}
\eqno(o)
$$
where $p^{in}(\Omega)$ and $p^{out}(\Omega)$ are the momentum distribution
at angle $\Omega=(\vartheta,\varphi)$ of the in - and out - going particles
respectively.
The shift function $f$, is the Green function defined on the event horizon
\refto{DT85,LS89} satisfying
$$
\nabla^2_{\perp}f-af=b\delta^2(\Omega-\Omega')~~,  \eqno(p)
$$
$$
a=2r_+\kappa (r_+)~~,~~~b=32\pi pr_+^2g_{uv}(r_+)~~.
$$
This expression can be integrated by using the properties of Legendre
polynomials\refto{GR80}
$$
f(\Omega-\Omega')=-{\pi b\over\sqrt{2}}{P_{-1/ 2+i{\sqrt{3}/ 2}a}
[-\cos(\Omega-\Omega')]\over \cosh{\left({\sqrt{3}\over 2}\pi a\right)}}~~.
\eqno(q)
$$
The Legendre functions $P_{-1/2+is}(z)$ are called conical functions
and are defined positives for $z<1$.

The poles of the S-matrix \(o) can be evaluated as follows. First we note that
the structure of this poles depends essentially on the short distance behavior
of the function $f(\Delta\Omega)$ (see ref. [\cite{LS92}]).
For our function \(q) this is given by
$$
f(\Delta\Omega)\approx {b\sin{(\pi\lambda)}\over\sqrt{2}
\cosh{\left({\sqrt{3}\over2}
\pi a\right)}}\left\{2\ln{\left({\Delta\Omega\over4}\right)}+\gamma+
\psi (\lambda+1)+\pi \cot{(\pi\lambda)}\right\}~~, \eqno(p1)
$$
where
$$
\lambda=-{1\over 2}+i{\sqrt{3}\over2}a~~;~~~\psi(z)={\Gamma'(z)\over \Gamma(z)}
{}~~~{\rm and}~~~\gamma=0.577215...
$$
What is essential here is the logarithmic dependence of $f(\Delta\Omega)$.
For this dependence the poles of the S-matrix are found to lie in\refto{T88}
$$
E^2_p=-ilM_{pl}^2~~;~~{\rm with}~~l~~{\rm a~positive~integer}~~.\eqno(p2)
$$
These imaginary poles give the bound state spectrum and correspond to the
ultrarelativistic hydrogen-like poles.

A functional integral representation can be given for the S - matrix\refto{T90}
$$
<p^{out}(\Omega)| p^{in}(\Omega)>={\cal N}\int{
{\cal D}x(\Omega) \exp{\left\{
\int{d^2\Omega\left[-{i\over 2\pi} \left((\partial_{\Omega}x)^2+
ax^2\right)+ixp^{ext}\right]}\right\}}}~~. \eqno(r)
$$
It is apparent the similarity between this expression and the partition
function
of our model, \eq{k}, if we identify there the two - dimensional
surface with the event horizon and the scalar order parameter with the
``membrane coordinates", $x$.

Near criticality the ``mass term", $\mu$ in \eq{k}, vanishes like
$\mu\sim |T-T_c|$~~.
Thus, the model becomes conformaly invariant. In this case we can write
the   functional integral formulation in a covariant ``Stringy" way
$$
{\cal Z}_G(J)={\cal N}\int{{\cal D}\phi^{\rho}(\sigma){\cal D}
g^{ab}(\sigma)\exp{\left\{-\int{d^2\sigma\left[-i\sqrt{g}g^{ab}\partial_a\phi^{
\rho}\partial_b\phi^{\rho}+i\phi^{\rho}J^{\rho}\right]}\right\}}}~~. \eqno(s)
$$
where $\sigma$ stands for the two horizon coordinates (in Euclidean space),
the order parameter has now $\rho $ internal dimensions and $g^{ab}$ is the
metric on the horizon surface.

Summarizing, we have started by showing the scaling of black holes near
criticality. This property of critical systems can be embodied in the
conformal invariance theory\refto{C87,I88}.
Then we are lead to string theory which is a realization of a conformal
field theory on the two - dimensional world - sheet\refto{GSW87}.
\eq{s} corresponds to the bosonic string case. The fermionic degrees of
freedoms can be eventually incorporated in it, this corresponding to the
addition of a fermionic order parameter in the effective hamiltonian \eq{m}.

Actually, for a continuum  model, the lower critical dimension is precisely
two\refto{B92}. This means that to have critical phenomena we should consider
$d=2+\epsilon$, where $\epsilon$ remains small whenever the hole be big
compared
to the Planck scale\refto{T85}.
Otherwise, if we consider a discrete model, the lower critical dimension
is one. We could thus keep $d=2$ and deal with a discrete order parameter
on the surface of black hole transformed now in a lattice with a spacing of
the order of the Plack length.

Of particular interest here is the result that the continuum limit of the
two - dimensional tricritical Ising model near the critical point is
supersymmetric\refto{Q86} (in ref. [\cite{L93}] we remarked the existence
of tricritical points in extreme Kerr - Newman black holes).

The results presented in this paper lead us to consider the following
effective model for dealing with a coarse grained quantization of black holes:

{\it A black hole appears to an external observer as if it had all its
quantum degrees of freedom concentrated on a thin membrane tightly
covering the horizon.}

This ``phenomenological"
model is of course observer - dependent, since in a reference system
falling with the matter that will form the black hole, nothing special
nor the membrane is seen when crossing the horizon.
It is also clear that it is the ``skin" on
the horizon that we propose to consider as a system to quantize by using
the standard rules of quantum field theory. In particular, we expect
an unitary S - matrix to exist, and to describe the process of formation and
evaporation of a black hole not leaving room for loss of quantum coherence.

\head{4. Discussion}

{}From the viewpoint of field theory the non - renormalizability of
quantum gravity is seen as a particularly annoying problem.
Especially since the establishment of the standard model of weak,
electromagnetic
and strong interactions, renormalizability has become a natural requirement
to any good theory. On the other hand, from the point of view of
statistical physics\refto{Z89} the non - renormalizability of gravity appears
natural, since its weakness suggests it is irrelevant (in critical phenomena
language) and therefore non - renormalizable. At low energies, far  below
$E_{pl}$, only the relevant operators  (which leads to renormalizable
theories) will survive. This explains why all current experimental
observations can be accurately described in terms of an effective long
distance gravitational theory. As $\beta$ - decay is the low - energy
remnant of much richer
physics above the electroweak scale,  New physics should be expected at
energies
$E>E_{pl}$. Our  effective model, presented in the last section,
can be thought of as the low energy version of the physics above $E_{pl}$
obtained by eliminating the details of its structure in a similar way as
is done, for example,  with the details of copper and zinc atoms from
the description of $\beta$ - brass to obtain the Ising model\refto{B92}.

We have shown conclusive evidence that black holes undergo critical
phenomena. Under this conditions their characteristic behavior is as if
they had an effective dimension equal to two (plus, eventually, $\epsilon$).
This dimensionality was first obtained by asking that the critical exponents
$\eta$ and $\nu$, derived from the correlation functions, satisfy the scaling
(and hyperscaling) relations. Then it was shown that by comparison of the
black hole other critical exponents with those of the Gaussian model in
$d$-dimensions, complete agreement can only be found for $d=2$. Finally, by
quite general arguments coming from the Renormalization Group theory, we
have argued that the effective dimension cannot be $d\geq3$.

The effective two - dimensionality (we remark that here this is not
imposed externally as in the case of 2 - $d$ black holes \refto{W91}) have
several interesting consequences.
Here we shall briefly discuss two of them.

A simple way to show  how the mass of a black hole should be quantized
can be obtained by describing the black hole by a wave function
corresponding to the order parameter in a critical system having one
component (a single scalar field) depending
only on the two angular coordinates that cover the horizon surface and the
proper time $t$.
In this simplified model the only effect of the black hole gravitational
field is to provide with the background geometry, i. e. the spherical
surface representing the horizon. If we impose to this wave function the
Klein - Gordon equation
$$
\left\{-\partial_t^2 +{1\over r_H^2}\nabla_{\Omega}^2 +\mu^2\right\}
\phi(\vartheta, \varphi, t)=0~~,  \eqno(D1)
$$
where $r_H=M+\sqrt{M^2-a^2-Q^2}$ is the horizon radius, we have the
following set of eigenfunctions
$$
\phi_{lm}=\exp{\left\{-iE_{l}t\right\}}Y_{lm}(\vartheta, \varphi)~~,
$$
where $Y_{lm}$ are the spherical harmonics and the energy of the system
is given by
$$
E^2_l=\mu^2+{\hbar^2 l(l+1)\over r_H^2}~~; ~~l=0, 1, 2,....\eqno(D2)
$$
Since $r_H\simeq {2GM\over c^3}$ (for a Schwarzschild hole)
and (if we consider that the whole black hole
is represented by\footnote{$^\#$}{For an alternative view see \ref{LC94}.}
$\phi$), $E_l\simeq M$, \eq{D2} implies
$$
M^2\simeq {\mu^2\over 2}+{1\over2}\sqrt{\mu^4+M_{pl}^4l(l+1)}~~.
$$
We have that for big $l$ (where we expect this approach to be valid)
$$
M\simeq {1\over\sqrt2}M_{pl}\sqrt{l}~~,~~{\rm with}~~l~~~
{\rm a~positive~integer}. \eqno(D3)
$$
This represents a quantization of the black hole mass in units of the
Planck mass. It is interesting to remark that the $\sqrt{l}$ dependence
have also been found in \eq{p2}, and by Bekenstein \refto{B74} using the
quantization of adiabatic invariant action integrals (see refs
[\cite{K86,M90,G93,P93,M94}] for still other independent derivations).
We note that \eq{D3} consist of three factors. While the first $1/\sqrt{2}$
term is expected to be model dependent, the $M_{pl}$ factor could have
been guessed on dimensional grounds. There seems to be some agreement in the
cited literature as well as in our \eqs{p2} and \(D3) on the $\sqrt{l}$
dependence. We thus think that \eq{D3} represents a first approximation
to the black hole mass quantization.
The black hole radiation will now come out in the form of a line spectrum
with most of the radiation at the frequency $\hbar\omega_l=\Delta Mc^2=
{M_{pl}^2c^2\over 4M}$ (also in multiples of this frequency), which corresponds
to the maximum of the (continuum) Hawking spectrum i.e. $\hbar\omega_{max}\sim
k_BT_{BH}\sim {M_{pl}^2c^2\over M}$.

Since our black hole system, as we have seen, has an associated effective
dimension equal to two, the proportionality entropy - area can be expected
to appear in a natural way.
In fact, since the mass of the black hole is quantized there must be a
finite number of internal states. They can be counted by noting \refto{M90}
that a black hole of mass $M$ given by \eq{D3} can be formed in $2^{l-1}$
different (and equivalent) ways from units of $M_{pl}$. The entropy
associated with the ignorance of the exact way in which the black hole formed,
can be evaluated, in a first approximation, as
$S_{bh}\simeq k_B\ln{[2^{l-1}]}~~.$ For large $l$ we have
$$
S_{bh}\simeq k_B\ln{2\left({M\over M_{pl}}\right)^2}
\simeq {k_Bl_{pl}\over 4\pi}\ln{2}\left( 4\pi r_H^2\right)~~, \eqno(D4)
$$
which gives the well - known proportionality between entropy and area of a
black hole.

One should not be bewildered by these results, since they are founded on
a crude approximation to the quantum black hole problem. The model is
necessarily incomplete (a second quantized description should be considered,
for example). Also 't Hooft suggests that the quantum states labeled
by $E_l$ in \eq{D2} are enormously degenerated\refto{FS91}. It is also
important to evaluate the width of each energy level (to account for
the quantum instability of black holes) and compare it to the separation
between energy levels\refto{P94}.
However, what we wanted to rescue from the above crude
model\footnote{$^\dagger$}{A more
detailed treatment is under study by the present author\refto{LC94}.} is
the relevance of the essentially two-dimensional nature of semiclassical
black holes and the possibility of representing them, in a first approximation,
by a single scalar field (playing the role of the order parameter in a
critical system).

Thus, in conclusion, we can say that black holes may have ``no hairs"
\refto{M82,B82}, but instead they seem to behave as if they had ``skin".

\vskip 30pt
\noindent
{\it Acknowledgements:}
\noindent
\mangos\dos

\vfill\eject

\references

\refis{K66} L. P. Kadanoff, Physics, {\bf 2}, 263 (1966).\par

\refis{HS72} A. Hankey and H. E. Stanley, {\it Phys. Rev. B}, {\bf 6}, 3515
 (1972).\par

\refis{F82} M. E. Fisher, in {\it Critical Phenomena}, F. J. W. Hahne Ed,
Springer Verlag, Berlin (1982), p 1-139.\par

\refis{SCD81} D. W. Sciama, P. Candelas and D. Deutsch,
{\it Advances in Physics}, {\bf 30}, 327 (1981).\par

\refis{A84} D. J. Amit, {\it Field Theory, the Renormalization Group, and
Critical Phenomena}, World Sci., Singapore, 2nd Ed. (1992).\par

\refis{B92} J. J. Binney, N. J. Dowrik, A. J. Fisher and M. E. J. Newman,
{\it The theory of Critical Phenomena}, Clarendon Press, Oxford (1992).\par

\refis{BK52} T. H. Berlin and M. Kac, {\it Phys. Rev. D}, {\bf 86},
821 (1952).\par

\refis{L93} C. O. Lousto, {\it Nucl. Phys. B} {\bf 410}, 155 (1993).\par

\refis{HH76} J. B. Hartle and S. W. Hawking,{\it Phys. Rev. D} {\bf
13}, 191 (1976).\par

\refis{T90} G. 't Hooft, {\it Nucl. Phys. B} {\bf 355}, 138 (1990).\par

\refis{DT85} T. Dray and G.'t Hooft, {\it Nucl. Phys. B} {\bf 253}, 173
(1985).\par

\refis{LS89} C. O. Lousto and N. S\'anchez, {\it Phys. Lett. B} {\bf 220},
 55 (1989).\par

\refis{I88} C. Itzykson, H.Saleur and J.B. Zuber, {\it Conformal invariance
and Applications to Statistical Mechanics}, World Sci., Singapore (1988).\par

\refis{LM93} R. M\"uller and  C. O. Lousto,
{\it Phys. Rev. D} {\bf49}, 1922 (1994).\par

\refis{GSW87} M. B. Green, J. H. Schwarz and E. Witten, {\it Supestring
Theory}, Vol. 1, Cambridge Univ. Press., Cambridge (1987).\par

\refis{T85} G. 't Hooft, {\it Nucl. Phys. B}, {\bf 256}, 727 (1985).\par

\refis{FS91} G. 't Hooft, {\it Physica Scripta}, {\bf T36}, 247 (1991).\par

\refis{Q86} Z. Qiu, {\it Nucl. Phys. B}, {\bf 270}, 205 (1986).\par

\refis{Z89} J. Zinn - Justin, {\it Quantum Field Theory and Critical
Phenomena}, Oxford Univ. Press, Oxford (1989).\par

\refis{C87} J. L. Cardy, in {\it Phase Transitions and Critical Phenomena},
Eds. C. Domb and J. L. Lebowitz, Vol. 11, Academic Press,
London, p 55 (1987).\par

\refis{1} J. M. Bardeen, B. Carter and S. W. Hawking, {\it Commun. Math.
Phys.}, {\bf 31}, 161 (1973).\par

\refis{18} K. G. Wilson, {\it  Rev. Mod.  Physics.}, {\bf 55}, 583 (1983).
\par

\refis{19} H. E. Stanley,  {\it  Introduction to Phase Transitions and
Critical Phenomena}, Oxford Univ. Press, Oxford (1971).\par

\refis{WK74} K. G. Wilson and J. Kogut, {\it Phys. Rep. C}, {\bf 12}, 75(1974).
\par

\refis{W92} R. M. Wald, in {\it Black Hole Physics},  V. De Sabbata and Z.
Zhang
 Eds, Nato Asi series, Vol. 364 (1992),  p 55-97. \par

\refis{9} J. D. Bekenstein,  {\it Phys. Rev. D}, {\bf 7}, 949 (1973); ibid,
{\bf D7}, 2333 (1973); ibid, {\bf D9}, 3292 (1974).\par

\refis{4} P. C. W. Davies, {\it Proc. R. Soc. Lond. A}, {\bf 353}, 499
 (1977).\par

\refis{26} D. Tranah and  P. T. Landsberg, {\it Collective Phenomena},
{\bf 3}, 81 (1980).\par

\refis{32} P. Hut, {\it Mon. N. R. Astr. S.}, {\bf 180}, 379 (1977).\par

\refis{TPM86} K. S. Thorne, R. H. Price and D. A. Macdonald (Eds.),
{\it Black holes: The membrane Paradigm}, Yale Univ. Press, New Heaven
(1986).\par

\refis{L94} C. O. Lousto, in {\it Cosmology and Particle Physics},
V. De Sabbata and H. Tso-Hsiu Eds., NATO ASI series Vol. C427,
p 183-192, Kluwer Academic Pubublishers, Dordrecht (1994).\par

\refis{TH92} G. 't Hooft, in {\it Black Hole Physics}, V. De Sabbata and
Z. Zhang Eds, Nato Asi series, Vol. 364 (1992),  p 381-402  .\par

\refis{W91} E. Witten, {\it Phys. Rev. D}, {\bf 44}, 314 (1991).\par

\refis{LC94} C. O. Lousto and M. Maggiore, work in progress.\par

\refis{M90} V. F. Mukhanov,  in {\it Complexity, Entropy and the Physics of
Information}, Ed. W. Zurek, Addison-Wesley, Redwood city, p 47 (1990).\par

\refis{B82} G. Bunting, Ph. D. thesis, Dept. of Math., Univ. of
New England, Armilade, Australia (1983).\par

\refis{M82} P.O. Mazur, {\it J.  Phys. A}, {\bf 15}, 3173  (1982).\par

\refis{LS92} C. O. Lousto and N. S\'anchez, {\it Phys. Rev. D}, {\bf 46},
4520 (1992).\par

\refis{T88} G. 't Hooft, {\it Nucl. Phys. B}, {\bf 304}, 867 (1988).\par

\refis{GR80} I. S. Grandshteyn and I. M. Ryzhik, {\it Tables of Integrals
Series, and Products}, Academic Press, N. Y. (1980). I thank H. Vucetich
for this remark.\par

\refis{B74} J. D. Bekenstein,  {\it Lett. Nuovo Cim.}, {\bf 11}, 467
(1974).\par

\refis{STU93} L. Susskind, L. Thorlacius and J. Uglum, {\it Phys. Rev. D},
{\bf 48}, 3743 (1993).\par

\refis{G93} J. Garc\'\i a - Bellido, preprint hepth/9302127.\par

\refis{P93} Y. Peleg, preprint hepth/9307057.\par

\refis{M94} M. Maggiore, preprint hepth/9401027.\par

\refis{K86} Y. I. Kogan, JETP Lett, {\bf 44}, 267 (1986).\par

\refis{P94} D. N. Page,  preprint hepth/9305040.\par

\endreferences
\end